\documentclass[aps,prb,twocolumn,amsmath,showpacs]{revtex4}
\usepackage{graphicx}

\newcommand{\bvec}[1]{{\bf #1}}

\newcommand{\bv}[1]{{\bf #1}}

\newcommand{\bra}[1]{\left<#1\right|}

\newcommand{\ket}[1]{\left|#1 \right>}

\newcommand{\EqLabel}[1] { \label{#1} }

\begin{document}

\title{The longitudinal conductance of mesoscopic Hall samples with
  arbitrary disorder and periodic modulations}

\author{Chenggang Zhou}

\affiliation {Department of Electrical Engineering, Princeton
University, Princeton NJ 08544,USA}

\author{Mona Berciu}

\affiliation{Department of Physics and Astronomy, University of
British Columbia, Vancouver, BC V6T 1Z1, Canada }

\date{\today}

\begin{abstract}
We use the Kubo-Landauer formalism to compute the longitudinal
(two-terminal) conductance of a two dimensional electron system placed
in a strong perpendicular magnetic field, and subjected to periodic
modulations and/or disorder potentials. The scattering problem is
recast as a set of inhomogeneous, coupled linear equations, allowing
us to find the transmission probabilities from a finite-size system
computation; the results are exact for non-interacting electrons. Our
method fully accounts for the effects of the disorder and the periodic
modulation, irrespective of their relative strength, as long as
Landau level mixing is negligible.  In particular, we focus on the
interplay between the effects of the periodic modulation and those of
the disorder.  This appears to be the
relevant regime to understand recent experiments [S. Melinte {\em et
al}, Phys. Rev. Lett. {\bf 92}, 036802 (2004)], and our
numerical results are in qualitative agreement with these experimental
results. The numerical techniques we develop can be generalized
straightforwardly to many-terminal geometries, as well as other
multi-channel scattering problems.
\end{abstract}
\pacs{73.43.Cd}

\maketitle

\section{Introduction}

A significant amount of research has been devoted to the study of the
effect of a periodic potential on a two dimensional electron system
(2DES) placed in large perpendicular magnetic fields.  On the theory
side, the so-called Hofstadter butterfly\cite{Hofst} -- the fractal
energy spectrum associated with the periodically modulated 2DES, in
the absence of disorder -- had been predicted and studied even before
Hofstadter's seminal paper.\cite{Wannier1,Dieter,Harper,Azbel} Later,
the transport properties of such systems were studied by St\v{r}eda,
MacDonald and others.\cite{ Streda1,Streda2,MacStreda,MacDonald} After
the discovery of the Integer Quantum Hall Effect (IQHE),\cite{IQHE}
experimental results started to become available.\cite{Wulf,Weiss}
Pfannkuche and Gerhardts put forward a detailed theoretical treatment
of transport properties, including disorder effects.\cite{Gerhardts}
Their theory is analogous to the self-consistent Born approximation
(SCBA),\cite{WulfMac} and it suggests that the splitting of one single
Landau band into several subbands by the periodic modulation can in
principle be observed from measurements of the longitudinal
conductance $\sigma_{xx}$. The effect of disorder on the fractal
structure was found to be similar to its effect on the Landau level
(LL) structure (responsible for the appearance of IQHE): subbands in
each Landau level are broadened by disorder, but energy gaps or pseudo
gaps are still open if the disorder is small compared to the amplitude
of the periodic potential. As the Fermi energy sweeps through a
subband, the longitudinal conductance has a maximum and the Hall
conductance shows a staircase-like jump if the Chern numbers in the
neighboring gaps are different. Thus, the Hall conductance is
expected to follow a nontrivial sequence of integer multiples of
$e^2/h$,\cite{denijs} whereas the longitudinal conductance has a
series of peaks and valleys as the Fermi level moves through
different subbands and gaps.

 A recent experiment\cite{Sorin} on a high-quality, periodically
modulated 2DES shows interesting new features in the longitudinal
conductance, although the periodic potential is too weak to produce
well-separated subbands (or, equivalently, disorder is strong enough
to fill in  all sub-gaps in the fractal structure of each Landau
level). Even in this case, the longitudinal conductance exhibits
reproducible oscillatory features in the presence of the weak periodic
modulation, instead of the single smooth Lorentz peak of the
unpatterned samples. To our knowledge, this regime of strong disorder
and weak periodic modulation has not been investigated in the
literature, and therefore these recent experimental results do require
theoretical interpretation. In recent work,\cite{Previous} we analyzed
the spectrum and nature (localized or extended) of electronic states
in such a regime, and showed that simple arguments based on these
results provide a qualitative explanation of the experimental
observations.

In this study, we present a numerical calculation of the longitudinal
conductance based on models appropriate for the type of samples used
in the experiment of Ref.~[\onlinecite{Sorin}].  Our model includes a
disorder potential and a periodic potential with either square or
triangular symmetry, with arbitrary relative strengths.  In this work
we assume that these potentials are small enough that Landau level
mixing is negligible, although the methods we develop can be trivially
generalized to take such mixing into account.  Unlike previous
theoretical studies dealing with disorder effects in QHE, which
performed average over disorder at the onset of the calculation so
that all computed response functions are disorder-averaged, we
calculate the longitudinal conductance from first principles for a
given disorder realization. This is necessary because the features
observed in this experiment\cite{Sorin} are believed to be
sample-specific. Our calculation is based on  the Kubo-Landauer
formalism.\cite{FisherLee} Our method is in principle valid for
finite systems with any type of disorder and/or periodic potentials,
although computational times vary with the sample size and degree of
sparseness of the Hamiltonian. Regrettably, we have no detailed
microscopic knowledge of the disorder present in these samples and the
details regarding their connections to the external leads; this
prevents us from performing meaningful quantitative comparisons with
the experiment. However, the results we obtain for different
realizations of disorder allow us to qualitatively explain the physics
responsible for the new features in the longitudinal conductance, and
to reinforce the arguments offered in our previous
work.\cite{Previous}

The paper is organized as follows: in Section \ref{sec2} we present
the method used for the calculation and the relevant theoretical
considerations. The numerical results are presented in Section
\ref{sec3}, while Section \ref{sec4} contains our conclusions and
discussions.

\section{The Numerical Method:  Kubo-Landauer formalism}
\label{sec2}

\subsection{The model}
\label{sec2.1}

We consider a two-dimensional Hall sample of rectangular shape, of
size $L_x\times L_y$, with cyclic boundary conditions in the
$y$-direction and open boundary condition in the $x$-direction,
characteristic of a two-terminal geometry. Typical sizes we consider
are on the order of 3$\mu m\times 3\mu m$. A large magnetic field
${\bf B}$ of up to 10T is applied in the $z$-direction, perpendicular
to the 2DES.  A rough estimate shows that the degeneracy $N$ of each
Landau level (LL) is of order $10^4$, defining the size of the matrix
to be diagonalized to be $10^4\times 10^4$.  For such large-size
matrices direct diagonalization is time-consuming, therefore  we look for alternative
approaches with a better scaling behavior for large systems. 
On the other hand, the sample size is
still small compared to that of the experimental sample, which is
about $20\mu m \times 20\mu m$. As a result, care must be taken in
interpreting the numerical results.  In principle it is possible to
increase the values of $L_x$ and $L_y$, however serious numerical
difficulties arise when the system size is much larger than the ones
we consider (these issues are discussed in section \ref{sec2.3}).

The Hamiltonian of the non-interacting electrons confined in the 2DES
is
$$ {\cal H}_s = \frac{1}{2m} \left(\bv{p}+\frac{e}{c}\bv{A}\right)^2 -
\frac{1}{2} g\mu_B\vec{\sigma}\cdot \bv{B} +V_d(x,y) + V_p(x,y).
$$ where $V_d$ and $V_p$ are the disorder and the periodic potentials,
respectively.  We use the Landau gauge $\bvec{A} = (0, Bx,0)$
throughout this paper, and the complete set of eigenfunctions for the
$n$\textsuperscript{th} Landau level:
\begin{equation}
\EqLabel{2.1} \langle \bvec r | n, X, \sigma \rangle = { e^{-i{Xy
\over l^2}} \over \sqrt{L_y}} e^{-{1 \over 2l^2}\left( x -X \right)^2
} { H_n\left({x -X \over l}\right) \over
\sqrt{2^nn!\sqrt{\pi}l}}\chi_\sigma
\end{equation}
where $l = \sqrt{ \hbar c \over eB}$ is the magnetic length and
$H_n(x)$ are Hermite polynomials.  In the rest of the paper, we
concentrate on one of the spin-polarized lowest Landau levels (LLL),
therefore we set $n = 0$. This is justified because in the
experiment both the disorder and the periodic potentials are estimated
to be much smaller than the cyclotron energy $\hbar \omega$ and the
Zeeman splitting,\cite{Sorin2} so that Landau level mixing can be
safely ignored.\cite{Sorin,Previous} Imposing cyclic boundary
condition in the $y$-direction leads to the restriction $X_j = j {2\pi
l^2 / L_y } $, $j=1,\cdots,N$. $X_j$, the guiding center, characterizes
the location at which  individual basis states are centered along
$x$-axis [see Eq.~(\ref{2.1})]. Since $X_j$ can vary between $0$ and
$L_x$, the degeneracy of each spin-polarized LLL is $N= L_xL_y /(2\pi
l^2)$.

Let us define $c^\dagger_j$ to be the creation operator for an
electron in the LLL: $c^\dagger_j|0 \rangle = |X_j\rangle$ (the
indexes $n=0$ and $\sigma$ will be suppressed from now on).  In the
absence of Landau level mixing, the Hamiltonian ${\cal H}_s$ projected
on the subspace of the spin-polarized LLL becomes:
\begin{equation}
\EqLabel{2.1a} {\cal H}_s = {\hbar \omega_c - g \mu_B B\sigma \over 2}
\sum_{j=1}^{N} c^\dagger_jc_j + \sum_{i=1}^{N}\sum_{j=1}^{N} \langle
X_i | V_d + V_p| X_j \rangle \cdot c^\dagger_ic_j
\end{equation}
This looks like a one-dimensional(1D) hopping Hamiltonian, and this is a
very appropriate comparison if one keeps in mind that the state
$|X_i\rangle$ is indeed localized within a distance $l$ of $X_i = i
2\pi l^2/L_y$.

In order to calculate the matrix elements for the disorder and the
periodic potentials, we use the identity
\begin{equation}
\EqLabel{2.3} 
\bra{ X_i} e^{i\bvec{q\cdot r}}\ket{X_j} = \delta_{X_i,X_j-q_yl^2}
e^{{i\over 2}q_x(X_i+X_j)}e^{-{1 \over 2} Q} 
\end{equation}
where $Q = {1 \over 2}l^2(q_x^2+q_y^2)$. (The generalization for higher
Landau levels and/or Landau level mixing is straightforward, see for
instance Ref. [\onlinecite{Gerhardts}]). Let us now consider each type
of potential separately.

The periodic potential can be expanded as:
\begin{equation}
\EqLabel{2.2} V_p(\bvec r) = \sum_{\bvec g} V_{\bvec g} e^{i\bvec{
r\cdot g}}.
\end{equation}
where $V_{\bvec g} = V_{-\bvec g}^*$ because $V(\bvec r)$ is real, and
$\left\{\bvec g\right\}$ are the reciprocal vectors associated with
the Bravais lattice. For a square potential, we use $V_{\bvec g}=A$
for all four shortest reciprocal vectors $\bvec g= (\pm {2\pi /a },0)\, ,
(0, \pm {2\pi /a })$, where $a$ is lattice constant, and zero
otherwise. Higher order components can also be included in the same
formalism, but result in longer computational time and no qualitative
changes.  Similarly, for a triangular potential we define $V_{\bvec g}
= -A$ for all six shortest reciprocal vectors ${\bvec g} = {4\pi /
\sqrt{3}a }(\pm 1,0)\, , {2\pi /\sqrt{3}a }(\pm 1, \pm \sqrt{3})$, and
zero otherwise. The minus sign appears here in order to have the
minima on the sites of the triangular lattice, as explained in
Ref. [\onlinecite{Previous}]. In both cases the projection of $\bvec
g$ on the $y$-axis is either $0$ or $\pm {2\pi \over a}$. This
particular orientation allows us to treat these two potentials
similarly, since it follows that both types of periodic potentials 
only couple a state $\ket{X_j}$ to itself and to $\ket{X_j \pm {2\pi
l^2 / a}}$ [see Eq.~(\ref{2.3})].  Since $\ket{X_j \pm {2\pi l^2 /
a}}$ must be in the basis considered, ${2\pi l^2 / a}$ must be an
integer multiple of $2\pi l^2 / L_y$, i.e. $L_y$ is an integer
multiple of $a$. This is consistent with the periodic boundary
conditions along the $y$-axis.

We introduce the integer $N_c = L_y/a$. From the previous discussion,
it follows that the periodic potential couples a state $X_j$ only to
itself, and to the states $X_{j\pm N_c}$. As a result, we can divide the
total $N$ states of the LLL into $N_c$ subclasses (henceforth called
the conduction channels) using the unique decomposition $j=iN_c+n$,
where $i>0$ is an integer, and $1 \le n \le N_c$.  The periodic
potential couples only states in the same channel $n$; these states
are distributed equidistantly across the sample, between the $x=0$ and
the $x=L_x$ edges, and can carry currents across the sample (hence the
name ``conduction channels''). For simplicity, we require that each
channel has the same overall number of states, i.e. the total
degeneracy $N$ is an integer multiple of the number of channels $N_c$,
$N=pN_c$. (This condition can be easily relaxed.) 
This imposes a constraint $L_x = p (2\pi l^2)/a$ on the values allowed for
$L_x$.
For the typical sample sizes we consider,
the constraints on $L_x$ and $L_y$ require only minimal adjustments. 
For instance, we
use $a=39$~nm,\cite{Sorin} and therefore a sample with $L_x\approx
L_y\approx 3~\mu$m will have around $N_c =70$ channels, with around
$p=150$ states per channel. For different values of the magnetic field
(different $l$ values) the length $L_x$ can be kept fixed within a few
nm by slightly adjusting the value of $p$.

For later convenience, we re-label the creation operators for states
in the LLL as $c^\dagger_j \rightarrow c^\dagger_{i,n}$, where
$j=iN_c+n$, $1\le n \le N_c$. It follows that the periodic potential
projected on the spin-polarized LLL takes the simple form
\begin{equation}
\EqLabel{2.1b} \hat{V}_p= \sum_{n=1}^{N_c} \left[\sum_{i=1}^{p}
\epsilon_{i,n} c^\dagger_{i,n} c_{i,n} +\sum_{i=1}^{p-1}
\left(t_{i,n}c^\dagger_{i,n} c_{i+1,n} + h.c. \right)\right],
\end{equation}
where for the square potential, we have:
\begin{subequations}
\EqLabel{2.15a}
\begin{eqnarray}
\epsilon_{i,n} &= &2A e^{- \left({\pi l \over a }\right)^2} \cos
\left[\left( i + {n \over N_c}\right) \left( { 2 \pi l \over
a}\right)^2 \right], \\ t_{i,n} &=& 2A e^{-\left({\pi l \over a
}\right)^2},
\end{eqnarray}
\end{subequations}
and for the triangular potential, we have:
\begin{subequations}
\EqLabel{2.15b}
\begin{eqnarray}
\epsilon_{i,n} = -2A e^{-{4 \over 3}\left( \pi l \over a\right)^2}
\cos\left[ { 8 \pi^2 l^2 \over \sqrt{3}a^2} \left( i + {n \over N_c}
\right)\right],&&\\ t_{i,n} = -2A e^{-{4 \over 3}\left( \pi l \over
a\right)^2} \cos\left[ { 4\pi^2 l^2 \over \sqrt{3} a^2} \left( i + {n
\over N_c}+{1 \over 2} \right) \right].&&
\end{eqnarray}
\end{subequations}

From Equations~(\ref{2.15a}) and (\ref{2.15b}), it is apparent that the
parameter controlling the band-structure (in the absence of disorder)
is the ratio $\phi/\phi_0 = {\cal A}/(2\pi l^2)$, where $ \phi_0=hc/e$
is the elementary flux, and $\phi = B {\cal A}$ is the magnetic flux
through the unit cell of the periodic potential. ${\cal A}=a^2$ or
$a^2 \sqrt{3}/2$ for square or triangular potentials respectively.  In
particular, if $\phi/\phi_0=q/p$, where $q$ and $p$ are mutually prime
integers, the original LL splits into $q$
sub-bands.\cite{Hofst,Wannier}

In the absence of disorder, there is no mixing between different
channels [see Eq.~(\ref{2.1b})], and the longitudinal current is just
a sum of the currents carried across the sample through  the individual conduction
channels. However, disorder introduces scattering between different
channels.  First principles modeling of the disorder in real samples
is a very difficult and numerically intensive
problem.\cite{Nixon,Davies,MStopa} As a result, we generate the
disorder potential using two simple phenomenological models
described in detail in Ref.~[\onlinecite{Previous}]. One is a simple
addition of random gaussians, while the second model attempts to
estimate the proper energy scale from considerations of the Coulomb
attraction between electrons and their donors.  Both models generate
smooth disorder potentials, i.e. with a length scale of more than
100~nm, which is very large compared with the typical magnetic length
$l$ ($l \sim 8$ nm when $B \sim 10$T). Such long-wavelength disorder
is believed to be dominant in high-quality samples, as the one studied
in Ref.~[\onlinecite{Sorin}].  The standard deviation is estimated to
be 2 to 3~meV, much smaller than the typical cyclotron
energy.\cite{Previous} We Fourier-decompose the disorder potential,
and use Eq.(\ref{2.3}) to compute its matrix elements.  The periodic
boundary condition in the $y$-direction implies that the allowed Fourier
components are $q_y = 2\pi m/L_y$, which introduces matrix elements
between any pair of states $|X_i\rangle$ and $|X_{i\pm m}\rangle$,
where $m$ is an arbitrary integer. It follows that different
conduction channels are now coupled by disorder.  The small $\bvec{q}$
Fourier components of the disorder potential are very important, since
they describe the long wavelength features of the disorder
potential. High values of $\bvec{q}$, on the other hand, describe
short wavelength features of the disorder, which are not well captured
by our simple phenomenological models. As a result, we use a cutoff
value of $|m|<36$ for Fourier components of the disorder potential
kept. This value is large enough to allow basically exact
reconstruction of the disorder potential (see relevant discussion in
Ref.~[\onlinecite{Previous}]) but also small enough so that the
Hamiltonian matrix is still very sparse. 
With this cutoff and
in the absence of LL mixing, the disorder potential has the general
form:
\begin{equation}
\EqLabel{2.1c} V_d = \sum_{n,n'=1}^{N_c} \sum_{i,i'=1}^{p}
v_{i,n;i',n'} c^\dagger_{i,n} c_{i',n'},
\end{equation}
where $v_{i,n;i',n'}$ is non-vanishing only for states within a
distance $|(i-i')N_c + (n-n')| \leq 36$ of each other.

From Equations~(\ref{2.1a}), (\ref{2.1b}) and (\ref{2.1c}) it follows that
the total Hamiltonian for the sample is:
$$ {\cal H}_s = \sum_{n=1}^{N_c} \left[\sum_{i=1}^{p} \epsilon_{i,n}
c^\dagger_{i,n} c_{i,n} +\sum_{i=1}^{p-1} \left(t_{i,n}c^\dagger_{i,n}
c_{i+1,n} + h.c. \right)\right]
$$
\begin{equation}
\EqLabel{2.1d} + \sum_{n,n'=1}^{N_c} \sum_{i,i'=1}^{p} v_{i,n;i',n'}
c^\dagger_{i,n} c_{i',n'},
\end{equation}
where the overall energy shift $(\hbar\omega_c- g\mu_B B \sigma)/2$
associated with the LLL is absorbed in a redefined chemical
potential. This Hamiltonian can be efficiently generated and stored as
a column compressed sparse matrix.  In principle, we can directly
compute the eigenvalues and eigenfunctions of this Hamiltonian, and
calculate the corresponding Thouless number, characterizing its
longitudinal conductance.\cite{Tnumber} However, this is numerically
very time-consuming.  Instead, we use the
Kubo-Landauer\cite{Landauer,Anderson80,Kubo,Baranger,FisherLee,Soukoulis}
formula for the longitudinal conductance (details in section
\ref{sec2.2}) which requires the computation of various transmission
coefficients through the sample. The main idea is to link the
longitudinal conductance to the total probability that an electron
injected into the sample at $x=0$ arrives at $x=L_x$, or vice versa.

\begin{figure}[tbp]
\includegraphics[width=0.9\columnwidth]{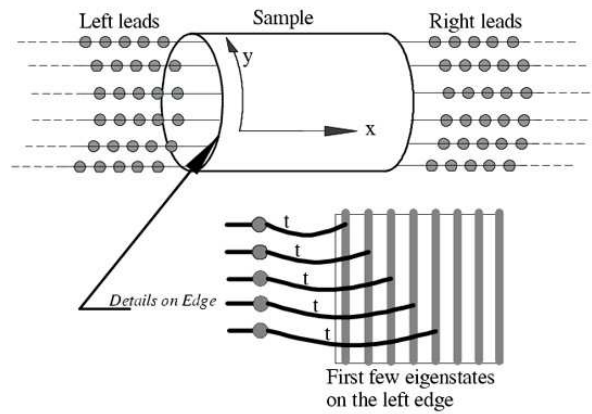}
  \caption{ A sketch of the model geometry of the Hall sample with
periodic boundary conditions in the $y$-direction, and its contact to
leads on both $x$-axis ends. The lower diagram shows a detailed view
of the left edge. The first $N_c$ eigenstates $|X_1\rangle,
|X_2\rangle ...,|X_{N_c}\rangle $ near the edge belong to different
conduction channels. We assume that each such conduction channel is
attached to external leads at both edges of the sample. }
  \label{fig:1}
\end{figure}

 In order to compute these transmission coefficients, it is necessary
to connect the sample to external metallic leads which allow us to
inject into and extract electrons from the sample.  We model each external
lead as a collection of independent, semi-infinite 1D
tight-binding chains, as illustrated in Figure \ref{fig:1}.  In reality,
the leads have, of course, higher dimensionality than one. One way to
simulate this would be to add bonds (hopping) between the various 1D
chains.  We do not add these extra bonds for the following reason: the
eigenstates of any lead with complex geometry in the transverse
direction have the general structure $E_{k,n}= \epsilon_n(k) +
u_n$. Here, $k$ is a quasimomentum associated with the longitudinal
direction, a good quantum number given the translational invariance
along this direction, and $n$ is some discrete set of
quantum numbers characterizing the discrete transverse modes supported
by the particular geometry of the lead cross-section. In other words,
any perfectly metallic higher-dimensional lead reduces to a collection
of independent 1D leads (or channels), whose dispersions
can be simulated by an appropriate choice of 1D tight-binding
chains.\cite{note5} The question, then, is how many channels are in
each lead, and how are they connected to the states in the
sample. Since we have no detailed knowledge regarding the leads, and
since one hopes that the main features of the longitudinal conductance
will come from the sample itself, not the details of the lead
modeling, we choose the following very simple solution: we assume that
both the left and right-side leads have precisely $N_c$ channels, and
each one of these channels couples identically to one of the
conduction channels inside the sample.  This is the simplest model
that satisfies several criteria: (i) the leads are perfect conductors;
(ii) the conductance of the leads is not less than the maximum
conductance of the sample;\cite{note} (iii) each conduction channel in
the sample has equal coupling to the leads; (iv) transmission and
reflection coefficients can be easily defined and computed. However, any other
more complex model for the connection of the sample to the external
leads can be investigated with the formalism we develop here.

Let us index with $n=1,2,...,N_c$ the $N_c$ independent 1D channels of
each semi-infinite lead, and let $c^\dagger_{i,n}$ be the creation
operator for an electron at site $i$ of the $n$\textsuperscript{th} 1D chain. We
distinguish between the left and right leads by requiring that $ i \le
0$ for the left channels, respectively $i > p$ for the right channels. This
notation also avoids any confusion between these operators, and the
operators creating electrons in LLL states inside the sample, which
correspond to $1\le i \le p$. The spin-index is suppressed everywhere,
since in the absence of magnetic impurities electrons with different
spins travel independently.
 
The Hamiltonian describing the external leads and their coupling to
the sample is then:
$$ {\cal H}_L = \sum_{n=1}^{N_c} \left(\sum_{i=-\infty}^{-1} +
\sum_{i=p+1}^{\infty}\right)\left[-t(c_{i,n}^\dagger c_{i+1,n} +
c_{i+1,n}c_{i,n}^\dagger) \right.
$$
$$ \left.+ \epsilon_0c_{i,n}^\dagger c_{i,n}\right] -t
\sum_{n=1}^{N_c}\left(c^\dagger_{0,n} c_{1,n} + c^\dagger_{p,n}
c_{p+1,n} + h.c.  \right)
$$ Here, $t$ and $\epsilon_0$ are adjustable parameters, whose
selection is discussed in section \ref{sec2.3}.  Our model of the
leads is thus similar to those employed for the study of molecular
conductance,\cite{Paulsson,Emberly,Onipko} where tight-binding chains
are used to simulate the gold contacts. In fact, we treat our Hall
sample as a big molecule wired to contacts. As we demonstrate in the
following, we only need to include explicitly 5 sites for each lead on
each side, because we apply outgoing or incoming boundary conditions
chosen so as to give the same result as explicit inclusion of
semi-infinite leads into our calculation.

The total Hamiltonian for the sample and leads reads:
$$ {\cal H} = \sum_{n=1}^{N_c} \sum_{i=-\infty}^{\infty}\left[\left(
t_{i,n} c_{i,n}^\dagger c_{i+1,n} + h.c.  \right) +
\epsilon_{i,n}c_{i,n}^\dagger c_{i,n} \right]
$$
\begin{equation}
+ \sum_{i,i'=1}^{p}\sum_{n,n'=1}^{N_c}v_{i,n;i', n'}c_{i,n}^\dagger
c_{i',n'}
\end{equation} 
where $t_{i,n} = -t$ and $\epsilon_{i,n}= \epsilon_0$ for all $i\le 0$
or $i \ge p$ (i.e., along the semi-infinite leads) whereas inside the
sample these parameters are given by Equations (\ref{2.15a}) and
(\ref{2.15b}).

\subsection{Formula for longitudinal conductance}
\label{sec2.2}

We calculate the longitudinal conductance using the formalism derived
in Ref.~[\onlinecite{FisherLee}], where it is shown that as a function
of Fermi energy $E_F$:
\begin{subequations}
\EqLabel{2.5}
\begin{eqnarray}
 \EqLabel{2.5a} \sigma_{xx}(E_F) &=& {e^2 \over h} \sum_{ab} v_av_b
 |G^{R,A}_{ba}(z,z', E_F/\hbar)|^2 \\ \EqLabel{2.5b} &=& { e^2 \over
 h} Tr(t^\dagger t),
\end{eqnarray}
\end{subequations}
Here, $z$ ($z'$) are positions in the asymptotic regions of the left
(right) leads, which are perfect conductors.\cite{Soukoulis} Also,
$v_{a(b)}$ are group velocities in channels a (b) of the corresponding
left (right) lead. The Green's function $G^{R,A}_{ba}(z,z',E_F/\hbar)$
for a continuous system is:\cite{FisherLee}
\begin{eqnarray}
\nonumber &&G^{R,A}_{ba}\left(z,z',{E_F\over\hbar}\right) = \\
  &&\int\int d\bvec{r}_\perp d\bvec{r'}_\perp
  u_a^*(\bvec{r'}_\perp)u_b(\bvec{r}_\perp)
  G^{R,A}\left(\bvec{r},\bvec{r'},{E_F\over\hbar} \right).
  \EqLabel{2.6}
\end{eqnarray}
The above equation is simply a projection of the full Green's function
$G^{R,A}(\bvec{r},\bvec{r'},E_F/\hbar)$ onto the transverse modes
$u_{a,b}$ of the leads.  In the asymptotic regions of the leads, the
conductance becomes independent of $z$ and $z'$ in Eq.~(\ref{2.5a})
because there is no dissipation in the leads.\cite{FisherLee} In our
model each lead is composed of independent 1D chains, each
representing a transverse mode (or channel). As a result, the
subscripts $a$ and $b$ in Eq.~(\ref{2.5a}) should be replaced by
indexes of different 1D chains on the left and right sides,
respectively. Eq.~(\ref{2.5b}) is the Landauer formula for
perfect-conductor leads,\cite{Landauer,Soukoulis} which we now
identify with Eq.~(\ref{2.5a}) by analyzing the structure of the
Green's function.

In Ref.~[\onlinecite{FisherLee}], Eq.(\ref{2.5a}) is derived for a
continuous model; the derivation can be replicated with minor
modifications for a discretized system such as ours.  The Green's
function is now labeled $G^R(i,n;j,m;E/\hbar)$, where $i$ and $j$
replace the locations $z,z'$ of the continuous model, whereas $n,m$
replace the channel indexes $a,b$.  Consider now the scattering state
$|\phi^+_{n,E}\rangle $ of an electron with momentum
$k>0$ and total energy $E(k) = \epsilon_0 - 2t\cos(ka)$ injected along
the $n$\textsuperscript{th} channel of the left lead. The scattering state is a solution of the
Lippman-Schwinger equation:
\begin{equation}
\EqLabel{2.7} |\phi^+_{n,E}\rangle = |\phi^s_{n,k}\rangle + { 1 \over
  E-{\cal H}_0+i0} ({\cal H} - {\cal H}_0)|\phi^+_{n,E}\rangle.
\end{equation}
Here, $|\phi^s_{n,k}\rangle$ is the incident wave along the left
chain, defined by $\langle j, m| \phi^s_{n,k}\rangle = \delta_{n,m}
e^{ikj}$, which is an eigenstate with energy $E $ of the Hamiltonian
${\cal H}_0$ defined by
$$ {\cal H}_0 = \sum_{n=1}^{N_c} \sum_{i=-\infty}^{\infty}
\left[\epsilon_0 c^\dagger_{i,n} c_{i,n} - t\left(c^\dagger_{i,n}
c_{i.n+1} +h.c. \right)\right].
$$ In other words, ${\cal H}_0$ describes $N_c$ non-interacting,
infinite 1D chains which go through the sample.  The scattering
Hamiltonian ${\cal H}-{\cal H}_0$ is then similar to the sample
Hamiltonian ${\cal H}_s$ [see Eq.~(\ref{2.1d})], except that
$\epsilon_{i,n} \rightarrow \epsilon_{i,n} - \epsilon_0$ and $t_{i,n}
\rightarrow t_{i,n} + t$ in order to account for the terms included in
${\cal H}_0$. As a result, the scattering potential ${\cal H}-{\cal
H}_0 $ has non-zero matrix elements only inside the sample.

The separation of the total Hamiltonian in this form is very
convenient, because the Green's function $G^R_0(E) = [E-{\cal H}_0 +
i0]^{-1}$ and corresponding group velocity $v_k$ for the tight-binding
model are well-known:
\begin{subequations}
\EqLabel{2.8}
\begin{eqnarray}
  \langle i, n| G^R_0( E) | j, m\rangle &=& \delta_{n,m} 
  {\exp(ik|j-i|) \over iv_k},\\ \EqLabel{2.8a} 
  v_k &=& \sqrt{4t^2 - (E-\epsilon_0)^2},\\ \EqLabel{2.8b} 
  E &=& \epsilon_0 -2t\cos(k). \EqLabel{2.8c}
\end{eqnarray}
\end{subequations}
The quasimomentum $k$ along the leads is measured in units of $1/d$,
$d$ being the lattice constant of the 1D chains.  (Since we assume
that all 1D chains are identical, the group velocity and dispersion
are the same for all of them. Generalization to non-identical channels
is straightforward, but requires inputs for more parameters).

One can now prove (see Appendix) that if $i<0$ and
$j>p$, then $G^R\left(i,n;j,m;{E\over\hbar}\right) =
\phi^+_{n,E}(j,m)e^{-iki}/(iv_k)$.  On any of the right-side channels, 
we have $\phi^+_{n,E}(j,m) = t_{nm}(E) e^{ikj}$ for all $j>p$, where
$t_{nm}(E)$ is the amplitude of probability of transmission from the
left channel $n$ into the right channel $m$ of an electron with energy
$E$.  Plugging this asymptotic form of $G^R$ into Eq.~(\ref{2.5a})
leads to the Landauer formula, Eq.~(\ref{2.5b}),
i.e. $\sigma_{xx}(E_F) = e^2/h \sum_{n,m}^{} |t_{nm}(E_F)|^2$.

The question, then, is how to efficiently calculate the transmission
coefficients. In principle, one needs to solve the Lippman-Schwinger
equation (\ref{2.7}) which, in turn, implies finding the retarded
Green's function $G^R(E)$ from the Dyson equation (see Appendix). The
essential difficulty for the numerical calculation is to contain
the infinite system in a finite-size scheme of computation with
appropriate boundary conditions.  Our solution to this problem 
allows us to find the scattering solution in a very elegant and economic way, 
which also avoids difficulties related to choosing the ``proper'' value for the
small imaginary parameter in the denominator.

To illustrate our solution, assume first, for simplicity, that there
is only one channel in both the left, and the right-side
leads. Consider the solution of the following inhomogeneous system of
linear equations:
\begin{widetext}
\begin{equation}
\EqLabel{2.9} \left(
\begin{array}[c]{ccccccccc}
e^{-ik} & -1 & 0 & 0 & \cdots &0 & 0 & 0 & 0 \\ t & E-\epsilon_0 & t &0 &
\cdots & 0 & 0 & 0 & 0 \\ 0& t & E -\epsilon_0& t & \cdots & 0 & 0 & 0 & 0
\\ \cdots &\cdots & \cdots& \cdots& \cdots& \cdots &\cdots & \cdots & \cdots \\ 0 & 0 & 0 & 0 &
\cdots& t & E-\epsilon_0 & t & 0\\ 0 & 0 & 0 & 0 & \cdots& 0& t &
E-\epsilon_0 & t \\ 0 & 0 & 0 & 0 & \cdots& 0 & 0 & -1 & e^{-ik} \\
\end{array}
\right) \left(
\begin{array}[c]{c}
\phi(-M+1) \\ \phi(-M+2) \\ \phi(-M+3) \\ \cdots  \\ \phi(p+M-2) \\ \phi(p+M-1)
\\ \phi(p+M) \\
\end{array}
\right) = \left(
\begin{array}[c]{c}
-e^{-ik(M-2)}\\ te^{-ik(M-1)}\\ 0 \\ \cdots \\ 0 \\ 0 \\ 0 \\
\end{array}
\right),
\end{equation}
\end{widetext}
where sites $-M+1,-M+2,\cdots, 0$ are on the left-side lead and sites
$p+1, \cdots , p+M$ are on the right-side lead, and $E = \epsilon_0 - 2t
\cos k$, with $k >0$.  The matrix in this equation is simply the
matrix $\langle i| E-{\cal H}|j \rangle$, dressed by the first and
last lines and columns. Along the lead sites the matrix has the simple
form of a hopping Hamiltonian, whereas inside the
sample, i.e. for sites between 1 and $p$, it contains the full
Hamiltonian of the sample.  One can now check, using iterations and
starting from site $-M+1$, that the general solution on the left-side
lead is of the form $ \phi(n) = e^{ikn} + \phi(-M+1) e^{-ik(M-1+n)}$ for
all $-M+1 < n < 0$. In other words, it consists of an incoming wave with
unit amplitude, and an outgoing wave with amplitude $|\phi(-M+1)|$. On
the other hand,  starting from the site $p+M$, one finds 
that the general solution on the right-side lead is of out-going type
$\phi(n) = \phi(M+p) e^{ik(n-M-p)}$ for all $n > p$. Since the matrix is
non-singular, this inhomogeneous system has a unique solution which
fixes the values of $\phi(-M+1)$ and $\phi(M+p)$. As a result, we
can identify directly the reflection coefficient for the given energy
$E$ as $r(E)=\phi(-M+1)$ and the transmission coefficient $t(E) =\phi(p+M)$.

This approach has a number of considerable advantages. First of all,
the number of sites $M$ kept on the left and right-side
leads is inconsequential, as long as it is greater than 2; the
solution obtained through the matching of the wavefunction inside the
sample is precisely the same as for semi-infinite leads. In our
simulations, we keep 5 sites for each 1D channel ($M=5$). The equivalence of the
solution of this finite system with that of a system with
semi-infinite leads also implies that while we work numerically with
finite matrices, we have a continuous energy spectrum $E(k) =\epsilon_0
- 2t \cos k$ for the electrons injected into the leads 
(generally, finite systems have a discrete spectrum).  Thirdly, there
is no need to introduce an infinitesimally small parameter $i0$ as in
Eq.~(\ref{2.7}), since the first and last line of the matrix insures
that it is no longer singular. Choosing the right value for the small
imaginary part is always difficult in numerical calculations and has
to be done very carefully in the case of a transfer
matrix.\cite{Kirkpatrick} With our approach, we take the limit $\eta
\rightarrow 0$ trivially and obtain the exact numerical
solution. Finally, one can use various entries on the right-hand side
of the inhomogeneous system. In our calculations, we actually use a
more efficient formulation, in which all coefficients on the
right-hand side, except the third entry, are zero, i.e. the equation
is $(E - {\cal H}) \phi = B$, where $B^T=(0 , 0, 1, 0, 0, \cdots.)$. One
can verify that this choice injects on the left lead, to the right of
the 3\textsuperscript{rd} site, an incoming wave with amplitude $1/(2it\sin
k)$ plus an outgoing wave, meaning that in this case, $t(E) = 2it
\sin k \cdot \phi(M+p)$. In this case (see Appendix) $\phi(n) = G^R(-M+1,
n, E)$, i.e. some of the matrix elements of the retarded Green's
function are produced.  Given the identity of Equations~(\ref{2.5a}) and
(\ref{2.5b}), the two formulations are equivalent, but the second is
numerically more efficient.

The generalization to the case with $N_c$ left and $N_c$ right
channels is straightforward. As already stated, we keep only 5 sites
for each of the left and right-side 1D channels, implying that $-4 \le
i \le p+5$. As a result the total dimension of the matrix $A = E-
{\cal H}$ is $N_A =pN_c +5\times2\times N_c$, where $N_cp = N$ is the
total number of states in the LLL. The matrix elements of $A$ equal
the values of $\langle i, n | E- {\cal H}| j, m\rangle$ for all $i,n$
and $j,m$ values, except for $\langle -4, n| A | - 4, n\rangle =
\langle p+5, n| A | p+5, n\rangle =\exp{(-ik)}$ and $\langle -4, n| A 
| - 3, n\rangle = \langle p+4, n| A | p+5, n\rangle =-1$ for all
$n=1,\cdots,N_c$. This insures that proper out-going solutions are
selected for each channel.

We then solve $n_0 =1,\cdots, N_c$ systems of inhomogeneous equations of
the type $\sum_{j,m}^{} A_{in, jm} X_{n_0}(j,m) = B_{n_0}(i,n)$, where
$B_{n_0}(i,n)= \delta(n-n_0) \delta(i+2)$, i. e. corresponds to an
electron injected into the $n_0$\textsuperscript{th} left-side channel. As discussed,
the transmission coefficients are then  $t_{n_0,m} = 2it \sin k
X_{n_0}(p+5,m)$ (one could choose any sites between $ p+2$ and $p+5$),
leading to the total longitudinal conductance at energy $E=\epsilon_0
- 2t \cos k$ to be
\begin{equation}
\EqLabel{2.11} \sigma_{xx}(E) ={ e^2 \over h} 4t^2\sin^2 k
\sum_{n_0=1}^{N_c} \sum_{m=1}^{N_c} |X_{n_0}(p+5,m)|^2
\end{equation}

 Despite having a very large dimension, the matrix $A$ is very sparse,
and the $N_c$ similar sets of linear equations $ AX=B$ for a given
energy $E$ can be solved very efficiently using the
SuperLU\cite{SuperLU} packages. This approach is much faster than
direct diagonalization and is particularly well suited for
parallelization; as a result, a dense grid of energy values $E$ can be
investigated. We used a cluster of 25 CPUs to scan different energy
values in parallel. A typical run lasts for about 10 hours and
generates 5000 data points of $\sigma_{xx}(E)$.

\subsection{ A toy model}
\label{sec2.3}

In this section, we analyze the longitudinal conductance for a simple
toy model. This allows us to understand the general effect of the lead
parameters $t$ and $\epsilon_0$ on longitudinal conductance $\sigma_{xx}(E)$. 
As shown in the following section, the shape of the
curves $\sigma_{xx}(E)$, especially for small or no disorder, is
rather surprising at first sight. It consists of large numbers of very
thin resonant peaks superimposed over a broad maximum. These sharp
resonance peaks are not numerical errors, and do not signal
singularities of the type expected in Green's functions.\cite{Previous}
(In fact, since we deal with an infinite system, one expects a
continuous cut, not individual singularities, in the Green's
functions.)  Zooming in the energy scale shows that these peaks are
features whose width scale inversely proportional to the size of the 
sample (i.e. region between the leads).   These resonance peaks have the same
origin as the peaks in the differential conductance of molecules
attached to metal contacts,\cite{Paulsson,Emberly,Onipko} i.e. they
correspond to resonant tunneling through the system. However, our 2DES
is different from a molecule in that it contains about $10^4$ internal
states, hence in a small energy interval there can be a large number of
such resonance peaks.

\begin{figure}[ht]
  \centering
  \includegraphics[width=0.9\columnwidth]{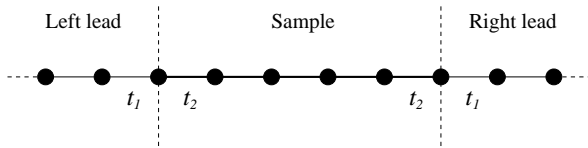}
  \caption{A sketch of the toy model. The ``sample'' with $n= 5$ $t_2$
  bonds is connected to semi-infinite leads with $t_1$ bonds.}
  \label{fig:1.5}
\end{figure}

A simple toy model that helps in clarifying the origin of these
resonance peaks is that of an $n$-bond tight-binding model with hopping
$t_2$ sandwiched between two semi-infinite tight-binding chains with
hopping $t_1$ (on-site energies are all zero for simplicity). 
Figure~\ref{fig:1.5} shows an example of the toy model.

We assume $t_2 < t_1$; transmission is then  vanishing for $|E|>2t_2$,
since the middle section (the ``sample'') does not support propagating
modes at those energies. The reflection and transmission coefficients on
the two interfaces (dashed line in Fig.~\ref{fig:1.5}) can be readily computed, 
and the total transmission rate is calculated either by summing up multiply
reflected waves, as for a Fabry-Perrot interferometer, or by solving
the Schr\"odinger's equation directly. The final result is summarized
below for a given energy $E$:
\begin{subequations}
\begin{eqnarray*}
E &=& -2t_1\cos(k_1) = -2t_2\cos(k_2)\\ x &=& t_1 \sin(k_1), \, y = t_2
\sin(k_2), \\ \alpha &=& x+y, \, \beta = x-y\\ T(E) &=& {16x^2y^2 
\over \alpha^4+\beta^4 - 2\alpha^2\beta^2\cos(2nk_2)}
\end{eqnarray*}
\end{subequations}

\begin{figure}[t]
\includegraphics[width=0.9\columnwidth,height=0.63\columnwidth]{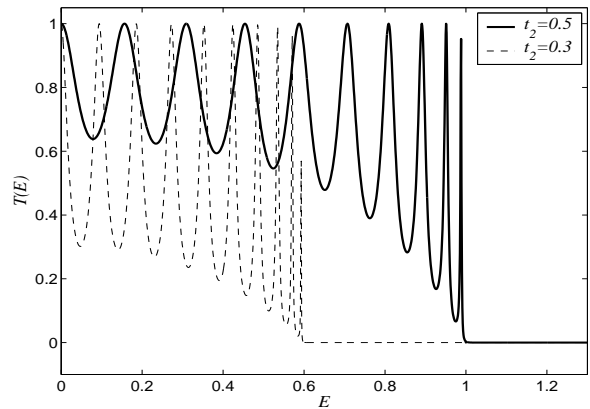}
\caption{Transmission rate for the toy model corresponding to $n=20$,
  $t_1=1$ and $t_2=0.5$ and 0.3 respectively. $T(E)$ is a symmetric
  function, so only the interval $E>0$  is shown. }
\label{fig:2}
\end{figure}

We can see from the last equation that $T(E)=1$ for all energies for
which $\cos(2nk_2)=1$, leading to the appearance of a series of thin
peaks at the corresponding energies.  The width of each resonance peak is
roughly $1/n$ of the band width, i.e. it is inversely proportional
to the system size, as indeed observed in Fig.~\ref{fig:2}. The minima
between neighboring peaks correspond to $T(E) = 4x^2 y^2/(x^2 +y^2)^2$, 
and are located at energies for which $\cos(2nk_2)=-1$.  
Thus, the larger the system, the more and sharper such resonance peaks appear. 
This toy model also provides a check for our numerical scheme, both in
the single and the multiple-channel cases.

The decrease in $T(E)$ from its maximum value of 1 is not due to the
resistivity of the ``sample'' itself, since there is no scattering
inside the sample in this toy model. Instead, it results from the
resistance of the contacts, which is due to the mismatch $t_1\ne
t_2$. (If $t_1 = t_2 \rightarrow \beta =0\rightarrow$, $T(E)=1$ at all
energies). In fact, if the ``sample'' strip is connected to left and
right leads in an infinitely smooth manner,  the conductance
simply becomes $e^2 /h$ within the overlap of the spectrum of the
``sample'' with the spectrum of the leads. 
Similar oscillating behavior in conductance due to contact
reflection was previously studied in Ref.~[\onlinecite{Kirczenow}],
where the conductance through a short and narrow ballistic channel is
calculated exactly. The oscillatory behavior is attributed to the {\it
``longitudinal resonant electron states, the electronic quantum analog
of the acoustic resonant modes of an open organ pipe''.} The toy model
we introduced above offers a good analogy to the continuous ballistic channel.

In our calculation, we want to concentrate on the physics inside the
sample, therefore we want to minimize the additional resistance
from the contacts, which introduces these extra features in the conductance
when disorder inside the sample is small. As a result, we have to 
adjust the value of $t$ so that it is close to the magnitude of the
 matrix elements $t_{i,n}$ of $H_s$. On the other hand,
the Fermi energy in the calculation is always required to be within
the spectrum of the leads,  $E_F \in [\epsilon_0 - 2t, \epsilon_0
  +2t]$, so that the leads behave like perfect 
conductors. To satisfy this condition without using a large $t$ value
(which leads to impedance mismatch), we set the on-site energy
$\epsilon_0$ at $E_F$ in each round of calculation, meaning that the
leads have a ``floating'' spectrum.  The floating spectrum is certainly 
not present in any experiments. We use it as a simple way to save computational power.
We have verified that such ``floating'' leads do 
behave as perfect metals in the simulation, and that small
variations of the parameters $\epsilon_0$ and $t$  do not change the
main features of the numerical results.

The appearance of the large number of resonance peaks also illustrates
the evolution from small quantum mechanical systems to
large macroscopic systems. Quantum mechanical quantities like
$\sigma(E_F)$ become rapidly oscillating functions as the size of the
system is increased. If the energy scale for such oscillations is
smaller than the resolution of the measurement or the temperature
smearing, the measurable physical quantity is a certain average of
this rapidly changing function within the characteristic energy
interval. In theory, numerical microscopic calculations become
inefficient for large systems, because one has to sample very many
energy values to obtain a good picture of the rapidly changing
function. In our case, we provide curves more suitable for comparison
with the experiment, by convoluting the zero-temperature conductance
with a sampling function:

\begin{subequations}
\EqLabel{2.13}
\begin{eqnarray}
        \EqLabel{2.13a} \bar\sigma_{xx}(\mu,V,T)   =
      \int_{-\infty}^{+\infty} \rho_T(\mu,V,\epsilon)
        \sigma_{xx}(\epsilon) d\epsilon, \hspace{5mm}&&\\
        \EqLabel{2.13b} \rho_T(\mu,V,\epsilon) = {1 \over V} \left[{1
        \over e^{\epsilon-\mu-V/2 \over kT}+1 } -{1 \over
        e^{\epsilon-\mu+V/2 \over kT}+1 } \right].&&
\end{eqnarray}
\end{subequations}
Here, $\bar\sigma$ is the measured conductance at chemical potential
$\mu$, temperature $T$ and voltage difference between the two edges of
the sample $V$ (estimated to be of order $10^{-6}$~eV). This formula
is appropriate for non-interacting electrons. One can verify that at
zero temperature,
\begin{equation*}
 \rho_0(\mu,V,\epsilon) = {1 \over V} [\Theta(\mu +V/2-\epsilon) -
        \Theta(\mu - V/2-\epsilon)].
\end{equation*}
When $V\rightarrow 0$, $\rho_T \rightarrow -{dn_{E_F}(T,\eta) / d\eta
  }$, where $n_{E_F}(T,\eta)=\left[\exp\left(\eta - E_F \over kT
  \right)+1\right]^{-1}$ is the Fermi-Dirac distribution.

\section{Numerical Results}
\label{sec3}

In this section, we present representative results from our
calculations. We analyze the interplay between disorder and periodic
potentials of various strengths, starting with the case of no disorder
(pure periodic potential).

\begin{figure}[t]
  \includegraphics[width=0.7\columnwidth,angle=-90]{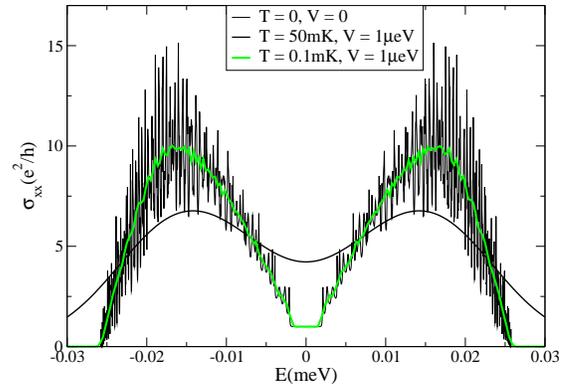}
  \caption{Longitudinal conductance of a small sample with square
  periodic potential but no disorder potential. $\phi/\phi_0=q/p =2$,
  $B$ = 5.44T, $L_x = 1.248\,\mu m$, $L_y = 1.17\,\mu m$,
  corresponding to a total of 1860 states in the lowest Landau level.}
  \label{fig:3}
\end{figure}

In Fig.~\ref{fig:3}, the longitudinal conductance of a small sample
with only square periodic potential is shown. There are only 1860
electronic states inside the sample, distributed over 30 channels and
the magnetic field is such that $\phi/\phi_0=q/p = 2$.  Because there
is no disorder, different conduction channels do not couple to each
other, and each of them is similar to the simple toy model discussed
in the last section. The spectrum of each conduction channel (except
one) splits into two subbands as expected for the Hofstadter butterfly
corresponding to $q = 2$. The channel that does not split corresponds
to $n=15$ in Eq.~(\ref{2.15a}), where the cosine vanishes and
$\epsilon_{i,n} = 0$ everywhere. Each channel contributes to the
conductance for energies inside its own spectrum, and thus we see many
sharp resonant peaks on top of a fairly broad conductance curve. As
already discussed, the sharp peaks are due to contact resistance, not
the sample itself. The underlying broad peaks,
on the other hand, are a signature of the sample behavior. They are
simply a reflection of the density of states inside the Hofstadter
butterfly in the clean model.  In the vicinity of $E=0$, we see that
$\sigma_{xx}$ is bounded by a series of staircases, each of which
marks the edge of the spectrum of a different conduction
channel. Thus, the curve for $T=0$ (thin black line) is understood
as a superposition of many 1D chains similar to the toy model shown in
Fig.~\ref{fig:2}.  The two solid curves in Fig.~\ref{fig:3} are the
``measured'' $\bar\sigma_{xx}$ given by Eq.~(\ref{2.13}), corresponding
to $V = 1\mu$eV and $T = 0.1$ and $ 50$~mK. At large temperatures, the
resonance peaks are smeared and only the two broad peaks are visible,
whereas for small temperature more detailed features are revealed.

\begin{figure}[b]
  \includegraphics[width=0.7\columnwidth,angle=-90]{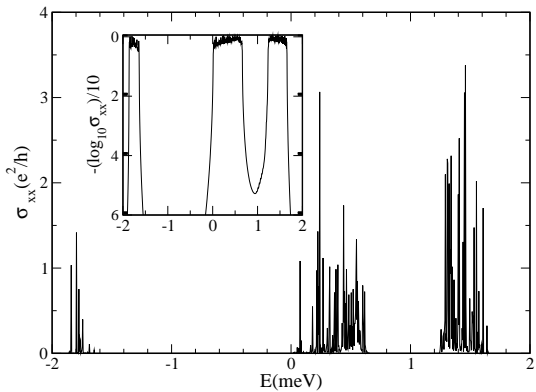}
  \caption{Longitudinal conductance for pure triangular periodic
  potential with $A=0.01$~meV (no disorder) and $q/p = 3$ ($B =
  9.42$T). $L_x=2.004\mu m,\, L_y = 2.028\mu m$. This sample has 9256 
  states in the LLL, divided into 52 channels.  The inset shows a
  semi-log plot of the original dataset.}
  \label{fig:4}
\end{figure}

\begin{figure}[t]
  \includegraphics[width=0.7\columnwidth,angle=-90]{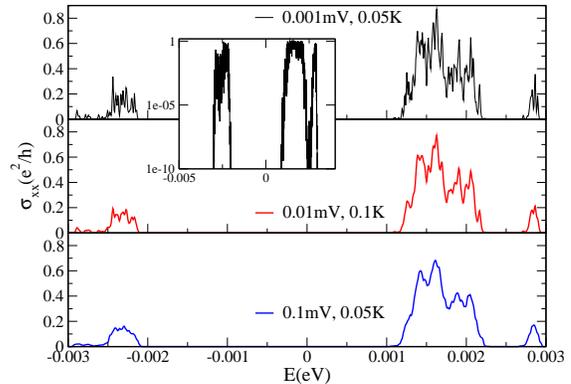}
  \caption{Longitudinal conductance for triangular periodic potential
  plus weak disorder, for $q/p = 7/3$ ($B$ = 7.33T). $L_x=1.998\mu
  m,\, L_y = 2.028\mu m$. This sample has 7176 states divided into 52
  channels. The inset shows a semi-log plot of the dataset. The
  amplitude of disorder is about 1meV, which is only a fraction of the
  size of the largest gap. However, some of the expected smaller gaps
  are already filled in by disorder.}
  \label{fig:5}
\end{figure}

Figure~\ref{fig:4} shows $\sigma_{xx}$ of a sample with only triangular
periodic potential at $q/p=3$.  From the semi-log inset, we can see
gaps between the $q=3$ expected adjacent subbands. Since the
triangular potential is not particle-hole symmetric (unlike  the
square potential) the subbands are no longer symmetrically placed
with respect to $E=0$.  In the smaller gap, the conductance is not
vanishing, although it is over 40 orders of magnitude smaller than
inside the band. This is a consequence of the fact that the sample is
finite and tunneling across it is possible even in the gap region 
(although with extremely low probability).

\begin{figure}[t]
  \includegraphics[width=0.8673\columnwidth]{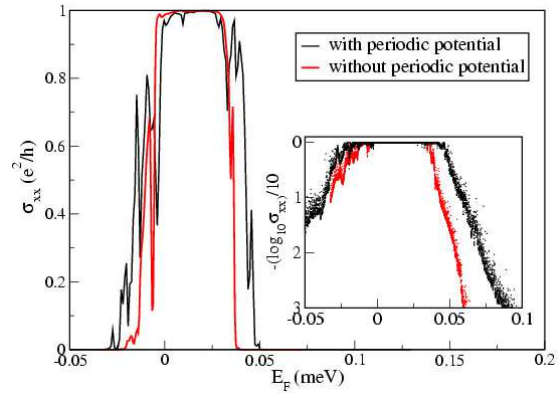} 
  \caption{Longitudinal conductance for a large disorder
  potential. $L_x=2.432\mu m$, $L_y=2.418\mu m$, $B=4.71$T. The thick
  line corresponds to disorder only, while the thin line corresponds
  to disorder plus a small periodic potential ($A=0.01$meV,
  $p/q=2/3$). In both cases $T=1mK$ and $V=1\mu eV$. The inset is a
  semi-log plot of the original datasets showing the increase of
  conductance in the off-resonance tunneling regime, induced by the
  small periodic potential. }
  \label{fig:6}
\end{figure}

Figure~\ref{fig:5} shows the calculated longitudinal conductance of a
sample with triangular potential and very small disorder, at
$q/p=7/3$. In the absence of disorder, we expect $q=7$ subbands to
appear asymmetrically in the spectrum. The small disorder closes some
of the gaps, and only three main subbands are distinguishable in
$\sigma_{xx}$. (Each of the three visible subbands is actually a
collection of 2 or 3 of the subbands expected in the absence of
disorder).  The three curves are the same data measured at different
temperature and different voltage drops across the sample.  As
expected, at low temperature and low voltage drop, the measured
conductance reveals a variety of resonance peaks on top of each broad
conductance feature. From the semi-log plot in the inset, one can see the
gap between the left subband and central subband is wide open, whereas
the smaller gap between the center and right subbands is partially
filled-in by disorder (the value of $\sigma_{xx}$ inside the gap is of
order $10^{-5}$, as opposed to $10^{-40}$ in the absence of disorder,
see Fig.~\ref{fig:4}).  This proves that narrow subbands are more
easily affected and therefore more likely to be closed by even small
disorder, as one would expect on general grounds. For higher
temperatures and/or voltage drops, the sharp resonance peaks are
averaged out and one obtains relatively smooth curves with broad peaks
reflecting the density of states and degree of localization of the
sample.

\begin{figure}[t]
\includegraphics[width=0.7\columnwidth]{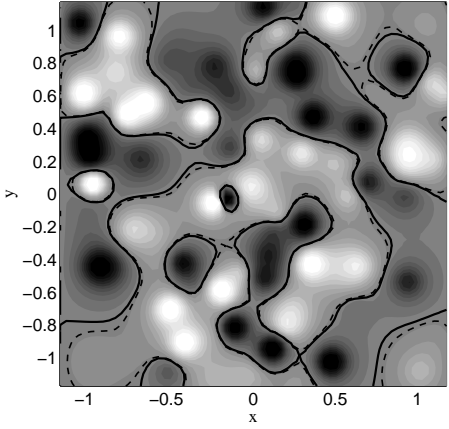}
\caption{Contour plot of the disorder potential used to calculate the
conductance of Fig.~\ref{fig:6}. Solid lines are equipotentials for
0~meV, and dashed lines for 0.025~meV. These energies are located
within the central peak of conductance. Parts of the contour go along
the edge at $y = \pm L_y/2$. }
\label{fig:7}
\end{figure}

In contrast to the previous cases, Fig.~\ref{fig:6} shows the
longitudinal conductance for a sample with only disorder potential
(thick line) and disorder plus a weak triangular periodic potential
(dashed line). In the disorder-only case we see a single broad peak
marking the conventional integer quantum Hall transition.  The curve
measured at $T=1mK$ and $V=1\nu eV$ is relatively smooth, and its flat
top at unit conductance indicates the existence of one semi-classical
orbit extending between the two opposite edges. This semi-classical
orbit can be seen in Fig.~\ref{fig:7}, where we display the disorder
potential used for this calculation. The conduction peak is not
centered at $E=0$ (center of the LLL level) because the disorder
potential is not fully particle-hole symmetric. However, 
if one symmetrizes the disorder and averages over
many disorder realizations, the averaged conductance peak would likely be a
smooth Lorentzian-type function. 
The finite (small) width is due to the finite size
of the sample: all wavefunctions with $x$-axis localization length
larger that $L_x$ can transport electrons between the leads.

\begin{figure}[b]
\includegraphics[width=0.9\columnwidth]{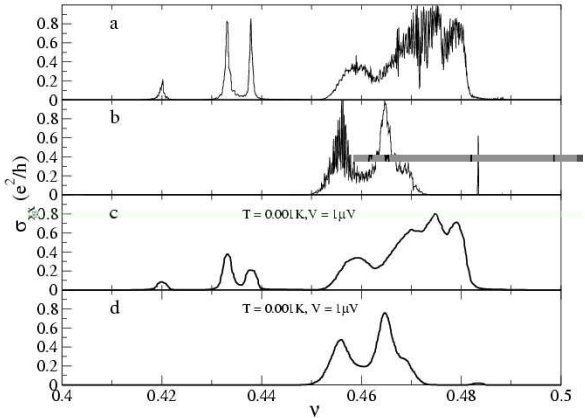}
\caption{Another example of a weak triangular potential imposed on a
large disorder potential. Here, $\sigma_{xx}$ is plotted as a function
of filling factor $\nu$. Panels (a) and (c) show the results with both
periodic and disorder potentials, panels (b) and (d) are for disorder 
only. Panels (a) and (b) are $T=V=0$ data, panels (c) and (d) are measured
at $T=1$~mK and $V=1\,\mu$V. Parameters are: $B=7.85~T$,
$p/q=2/5$, $L_x=2.432\mu$m, $L_y=2.418\mu$m, 11036 states divided in  62
channels. }
\label{fig:10}
\end{figure}

As expected, the addition of a weak periodic potential is not enough
to open gaps between the $q=3$ subbands of the corresponding
Hofstadter butterfly; the amplitude of the periodic potential is
only a small fraction of the bandwidth of the disorder-broadened Landau
level. However, the small periodic potential still has a sizable
effect on the longitudinal conductance: the magnitude of the off-peak
conductance has clearly been increased (see the inset) especially on
the high energy side. Also, the width of the central peak is increased
by an amount comparable to the magnitude of the periodic potential
$A$, with several more peaks separated by clear valleys appearing on
both sides of the central peak.

\begin{figure}[t]
\includegraphics[width=0.7\columnwidth,angle=-90]{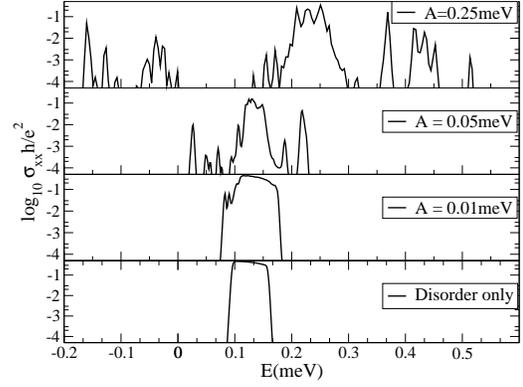}
\caption{Longitudinal conductance of a sample with a fixed disorder
  potential but varying strength of triangular periodic potential.
  Parameters for this sample are: $L_x = 3.107\mu m$, $L_y=2.964\mu
  m$, $p/q = 2/5$, $B = 7.85$T, leading to 16872 states divided
  amongst 76 channels. The data corresponds to $T=1$mK and
  $V=1\mu$eV. }
\label{fig:8}
\end{figure}

Figure~\ref{fig:10} shows another example of the effect of a weak
periodic potential on $\sigma_{xx}$. In this case, we plot
$\sigma_{xx}$ against the filling factor $\nu(E)$ calculated
as described in Ref.~[\onlinecite{Previous}]. For disorder only 
[panel (b) and (d)], we see a narrow, double-peaked conductance near half-filling,
in agreement with the general expectation for the IQHE (the
double-peak structure is an artifact of the particular disorder
realization used in this simulation). When a small periodic modulation
is added [panel (a) and (c)] the conductance shows a much more complex
shape: the central peak is broadened considerably, and several extra
peaks appear on the low-filling side.
 
\begin{figure}
\includegraphics[width=0.8\columnwidth]{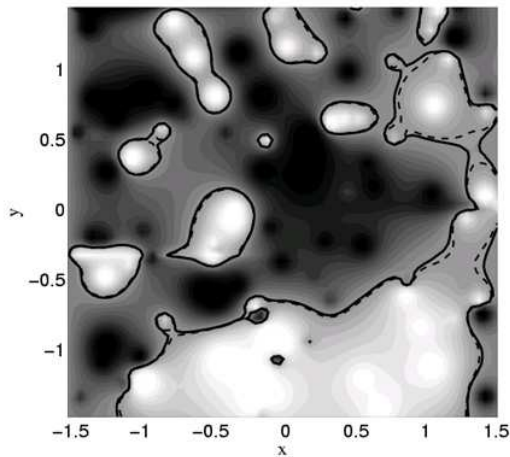}
\caption{The disorder potential used in the calculation for
Fig.~\ref{fig:8}, generated by summing random Gaussian
scatterers. Solid and dashed contours are at energies 0.11~meV and
0.14~meV respectively, marking the extended orbit responsible for the
conductance peak of the pure disorder case. }
\label{fig:9}
\end{figure}

Finally, Fig.~\ref{fig:8} shows the conductance of a disordered sample
for varying strengths of the periodic potential. The disorder
potential, which is kept fixed, is plotted in Fig.~\ref{fig:9}. For a
strong periodic modulation as compared to the disorder ($A=0.25$~meV,
upper panel), the $q=5$ subbands expected in the Hofstadter butterfly
at this magnetic field are beginning to emerge. On the other hand, in
the disorder-only case (lowest panel)  we see a single, smooth conductance peak
corresponding to the small energy interval where percolations through
the sample is established. Our results show that the interpolation
between the two cases  shows interesting and non-trivial behavior.
Although when $A=0.05$~meV the disorder is large enough to
completely erase the butterfly structure, the weak periodic modulation
has a non-trivial signature reflected by large numbers of peaks on
both sides of the disorder-only main conductance peak. The physical
origin of these extra peaks has been carefully analyzed in
Ref.~[\onlinecite{Previous}], where we argued that even a weak
periodic potential can efficiently create supplementary percolation
paths through the sample for energies within a range of the order $A$
from the critical region.  Here, we see the signatures of these states
in the longitudinal conductance as well-defined peaks within narrow
intervals in energy. Even an extremely small periodic potential (such
as $A=0.01$~meV) still widens the conductance peak, although there are
fewer extra peaks visible in this case.

One technical note: we used a semi-log scale in Fig.~\ref{fig:8}
 because the effect of the periodic potential is not so well revealed
 in linear scale. This is partially due to the fact that we kept $t$ for the
 leads constant for all four cases; this implies that as $A$ is
 changed, the impedance mismatch increases and the contact resistance
 becomes more and more important, suppressing the value of
 $\sigma_{xx}$.

\section{Conclusion and Discussions}
\label{sec4}

In this paper we use the Kubo-Landauer formalism to compute directly
the longitudinal conductance of a 2DES in the presence of periodic
modulations as well as disorder. The method employed allows us to
study individual disorder realizations and thus to analyze
sample-dependent effects. Our method treats the sample as a ``big
molecule'' connected between leads, and is applicable for any type of
one-electron potentials, both disordered and periodic. This
formulation is particularly suitable for long lengthscale, smooth
disorder potentials such as are believed to be dominant in high
quality 2DES, because in this case the Hamiltonian is a sparse matrix
that can be handled numerically very efficiently.  Our simulations are
performed for large mesoscopic samples (several microns in linear size),
corresponding to roughly $10^4$ electron states per Landau
level. Although this size is still smaller than that of most devices
used in IQHE experiments, this type of calculation can help us
understand the physical processes in detail. This method can also be
very fruitful for investigating transport in various nano-scale devices, and
can also be generalized in a straightforward manner for systems
connected to more than two terminals (leads).

Here, we concentrate on the interplay between a short wavelength
periodic potential and long wavelength disorder, and their role in
determining the longitudinal conductance of the 2DES. The
phenomenology in the asymptotic limits has been known for a long time.
A pure periodic potential splits the Landau levels into a number of
subbands, and the resulting band-structure has fractal properties as a
function of the magnetic field. In this case, our simulations show the
clearly separated peaks in longitudinal conductance corresponding to
each subband. If a weak disorder potential is added, the smaller gaps
are closed by disorder, as expected on general grounds.  Our results
are in good qualitative agreement with previous studies of the
weak-disorder case, using SCBA and similar approaches.\cite{WulfMac,Gerhardts} 
However, unlike our method, the SCBA
is valid only for small disorder and gives only disorder-averaged
quantities.  On the other hand, our method also reproduces the results
expected for large disorder (no periodic potential) with
long lengthscale, smooth variation. In this case we find a single,
relatively smooth peak in the conductance at energies where
semi-classical percolated orbits connect the opposite edges, and
negligible conductance elsewhere.

The new results come when we investigate the cases where the periodic
modulation is comparable, but smaller than the disorder. To our
knowledge, this case had not been investigated previously. We find
that even a weak modulation has a non-trivial effect on the
conductance, with the conduction being significantly enhanced. New
sharp peaks in conductance develop around the original conductance
peak, increasing its width and creating complex oscillatory features.
Although these peaks are clearly due to the periodic modulation, their
origin is not simply the Hofstadter structure, which is not relevant
for large disorder.  The mechanism of enhanced conduction was
explained in detail by us in
Refs.~[\onlinecite{Sorin,Previous}]. Basically, the periodic potential
helps electrons percolate through flat regions in the disorder
landscape and thus connect localized states to form new conductive
states. This is in qualitative agreement with recent experimental
observations\cite{Sorin} which show 
distinct patterns of peaks and valleys in the longitudinal conductance
of a periodically modulated 2DES, 
instead of the smooth peak expected for unpatterned samples.  However, our
calculation is limited to the {\it band conductance}, which has
metallic temperature dependence, and therefore is not suitable to
explain the temperature-enhanced conduction in the tail of $R_{xx}$
seen in the same experiment. We believe that in that region, the
conduction is due to hopping among localized states rather than charge
transport through extended states. As demonstrated in
Ref.~[\onlinecite{Previous}], a small periodic potential is very
effective in increasing the localization lengths and thus the hopping
probability.

Our method of computation has a number of limitations. One comes from
the fact that it is an exact calculation at zero temperature for a
large system.  Especially in the absence of disorder, $\sigma_{xx}$ is
a rapidly oscillating function that requires evaluation at a huge
number of energy values in order to get a good sampling for
temperature averaging, and this is a major limitation. Secondly, the
lead modeling is very elementary, although we 
can treat the case of semi-infinite leads exactly.  The simplicity in
describing the leads and their coupling to the sample comes from our
ignorance of their detailed physical characteristics. However, one can
straightforwardly generalize our method to describe more complicated
dispersion relations, and different types of channels and/or couplings
to the sample. In particular, the contact resistance due to the
mismatch between the leads and the sample should be minimized as much
as possible; this can be achieved with
a suitable choice of the lead parameters.

The most obvious limitation of this study is that we have little
knowledge of the functional form of the disorder and the periodic
potential, and how the sample's LL wave functions are coupled to the
Fermi sea in the leads. Until such knowledge is available, detailed
quantitative comparisons with the experiments are not warranted. We
use two simple phenomenological disorder models, and both give
qualitatively similar results. Coupled with our understanding of the
nature of the wave-function (see Ref.~[\onlinecite{Previous}]) this
gives us the confidence to claim that such effects are genuine and
should be observed for smooth, long length scale disorder. If the
disorder varies on a much shorter length scale, as may be the case for
quantum wires and dots (e.g., in Ref.~[\onlinecite{MStopa, Nixon}]),
we believe that a weak periodic modulation will have a very small, if
any effect, simply because in this case there are no relatively flat
regions in the disorder landscape where the periodic potential plays
a dominant role.  This is indeed confirmed by simulations we
performed for models with short length scale disorder (not shown here), where
addition of  a weak periodic modulations has no noticeable effect on the
longitudinal conductance.

 As for the periodic potential, we have been using only the components
with shortest reciprocal lattice vectors. However, as experimentalists
are designing new devices with enhanced periodic potentials, the
higher order Fourier components, as well as inter-Landau band mixing
which has not been considered here, might play a role. Such cases can
be treated with this formalism, the complications being only of
numerical nature. As more accurate models for the sample disorder and
modulation, as well as the leads and their coupling, on one hand, and
more powerful computational facilities, on the other hand, become
available, this formalism will allow for meaningful comparisons with
experimental results for transport in mesoscopic systems.

\section*{Acknowledgements}

We thank Sorin Melinte, Mansour Shayegan, Paul M. Chaikin and Mingshaw
W. Wu for valuable discussions.  This research was supported by NSF
grant DMR-0213706 (C.Z.) and NSERC of Canada (M.B.).

\appendix*
\section{Scattering States}

In the following we demonstrate the relationship between
 $G^R(-2,n_0;j,m; {E \over \hbar})$ and $\phi^+_{n_0,E}(j,m)$.  The
 scattering state $\phi^+_{n_0,E}(j,m)= \langle j,m|
 \phi^+_{n_0,E}\rangle$ satisfies the Lippman-Schwinger equation
 Eq.~(\ref{2.7}) for $n=n_0$, while the retarded Green's function
 $G^R(-2,n_0,j,m,{E \over \hbar})$ satisfies a similar Dyson equation:
\begin{equation}
\EqLabel{A2} G^R = G^R_0 + { 1 \over E - {\cal H}_0+i0 }( {\cal H
  -H}_0 ) G^R
\end{equation}
where, for simplicity, we do not write the arguments of $G^R$ and
$G^R_0$ explicitly. Here $(E - {\cal H}_0+i0 )^{-1}=G^R_0(E)$ is just
the formal expression in matrix form, and $i0$ is an infinitesimally
small imaginary number which selects out-going waves in the analytic
calculation. The inhomogeneous term in the Dyson's equation has the
form given by Eq.~(\ref{2.8}):
\begin{equation}
\EqLabel{A3} G^R_0(-2,n_0,j,m,{E\over \hbar}) = \delta_{n_0,m}{
\exp\left(ik|j+2|\right) \over iv_k}
\end{equation}
We multiply Eq.~ (\ref{2.7}) by $e^{2ik}\over iv_k$ and subtract it
from Dyson's equation Eq.~(\ref{A3}), to obtain:
$$ \psi = \delta_{n_0,m} \theta(-2-j) { \exp\left[-ik(2+j)\right]-
\exp\left[ik(2+j)\right]\over iv_k}
$$
\begin{equation}
\EqLabel{A4} +{ 1 \over E - {\cal H}_0+i0 }( {\cal H -H}_0 ) \psi
\end{equation}
where $\psi = G^R(-2,n_0,j,m, {E \over \hbar}) - { e^{2ik}
\phi^+_{n,E}(j,m) \over iv_k }$. This equation can be solved by
noticing that ${\cal H - H}_0$ has the same structure as ${\cal H}_s$,
whereas the inhomogeneous term has support only on the $n_0$\textsuperscript{th}
lead, outside the sample. It follows that the first order solution
$\psi =- \delta_{n_0,m} \theta(-2-j) { 2 \sin\left[k(2+j)\right] \over
v_k}$ is actually the exact solution, and therefore:
$$ G^R(-2,n_0,j,m, {E \over \hbar}) = { e^{2ik} \phi^+_{n,E}(j,m)
\over iv_k }
$$
\begin{equation}
- \delta_{n_0,m} \theta(-2-j) { 2 \sin\left[k(2+j)\right] \over v_k}
\end{equation}
This shows that up to a multiplicative constant, the Green's function
reproduces the scattering state on the right-side leads ($j>p)$, and
its asymptotic form there is therefore ${e^{2ik} \over iv_k} t_{n_0,m}
e^{ikj}$.


\begin{thebibliography}{99}


\bibitem{Hofst} D. R. Hofstadter, Phys. Rev. B {\bf 14}, 2239 (1976).

\bibitem{Wannier1}G. H. Wannier, Rev. Mod. Phys. {\bf 34}, 645 (1962).

\bibitem{Dieter} Dieter Langbein, Phys. Rev. {\bf 180}, 633 (1969).

\bibitem{Harper} P. G. Harper, Proc. Phys. Soc. Lond. A 68, 874
  (1955).

\bibitem{Azbel} M. Ya. Azbel', Zh. Eksp. Teor. Fiz. 46, 939 (1964);
  [Sov. Phys.-JETP 19, 634 (1964)].

\bibitem{Streda1} P. St\^{r}eda, J. Phys. C: Solid State Phys. {\bf
15}, L717 (1982).

\bibitem{Streda2} P. St\^{r}eda, J. Phys. C: Solid State Phys. {\bf
15}, L1299 (1982).

\bibitem{MacStreda} A. H. MacDonald and P. St\^{r}eda, Phys. Rev. B
{\bf 29}, 1616 (1984).

\bibitem{MacDonald} A. H. MacDonald, Phys. Rev. B. {\bf 29}, 6563
  (1984).

\bibitem{IQHE} K. von Klitzing, G. Dorda and M. Pepper,
Phys. Rev. Lett. {\bf 45}, 494 (1980).

\bibitem{Wulf} Rolf R. Gerhardts, Dieter Weiss, and Ulrich Wulf,
Phys. Rev. B {\bf 43}, 5192 (1991).

\bibitem{Weiss} D. Weiss, M. L. Roukes, A. Menschig, P. Grambow,
 K. von Klitzing and G. Weimann, Phys. Rev. Lett. {\bf 66}, 2790
 (1991).

\bibitem{Gerhardts} Daniela Pfannkuche and Rolf R. Gerhardts,
Phys. Rev. B {\bf 46}, 12606 (1992).

\bibitem{WulfMac} Ulrich Wulf and A. H. MacDonald, Phys. Rev. B {\bf
47}, 6566 (1993).

\bibitem{denijs} see, for instance, the non-trivial Chern numbers for
  the square lattice.

\bibitem{Sorin} S. Melinte, M. Berciu, C. Zhou, E. Tutuc,
S. J. Papadakis, C. Harrison, E. P. De Poortere, M. Wu, P. M. Chaikin,
M. Shayegan, R. N. Bhatt, and R. A. Register, Phys. Rev. Lett. {\bf 92},
036802 (2004).


\bibitem{Previous} Chenggang Zhou, Mona Berciu and R. N. Bhatt,
  cond-mat/0401007.

\bibitem{Sorin2} S. Melinte, E. Grivei, V. Bayot, and M. Shayegan,
Phys. Rev. Lett. {\bf 82}, 2764 (1999), and references therein.

\bibitem{Wannier} F. H. Claro and G. H. Wannier, Phys. Rev. B {\bf
19}, 6068 (1979).


\bibitem{Nixon} J. A. Nixon and J. H. Davies, Phys. Rev. B {\bf 41},
7929 (1990).

\bibitem{Davies} John A. Nixon, John H. Davies and Harold U. Baranger,
Phys. Rev. B {\bf 43}, 12638 (1991).

\bibitem{MStopa} M. Stopa, Y. Aoyagi, Physica {\bf B 227}, 61 (1996);
M. Stopa, Phys. Rev. B {\bf 54}, 13767 (1996); M. Stopa, Phys. Rev. B
{\bf 53}, 9595 (1996).

\bibitem{Tnumber} D. J. Thouless, Phys. Rep. {\bf 13}C, 93 (1974).

\bibitem{Landauer} R. Landauer, Philos. Mag. {\bf 21}, 863 (1970).

\bibitem{Anderson80} P. W. Anderson, D. J. Thouless, E. Abrahams, and D. S. Fisher, Phys. Rev. B {\bf 22}, 3519 (1980).

\bibitem{Kubo} Various equivalent forms of Kubo formula for
conductivity exist, see for example, R. Kubo, J. Phys. Soc. Jpn. {\bf
12}, 570 (1957) or R. Kubo, M. Toda and N. Hashitsume, {\em
``Nonequilibrium Statistical Mechanics''} (Springer, 1985), also
Refs.~[\onlinecite{Streda1,Streda2}]. Ref.~[\onlinecite{Baranger}]
offers detailed derivation of different forms, and
Ref.~[\onlinecite{FisherLee}] proves that in a certain type of model
(similar to ours), the Kubo and the Landauer formalisms are
equivalent.

\bibitem{Baranger}H. U. Baranger and A. D. Stone,
Phys. Rev. B {\bf 40}, 8169 (1989).

\bibitem{FisherLee} The Kubo formula and the Landauer formula have
been found to be equivalent in the current context for longitudinal
conductance, see Daniel S. Fisher and Patrick A. Lee, Phys. Rev. B
{\bf 23}, 6851 (1981).

\bibitem{Soukoulis} E. N. Economou and C. M. Soukoulis,
Phys. Rev. Lett. {\bf 46}, 618 (1981).

\bibitem{note5} As we introduce the computational method, it will also
  become clear that in the Landauer formula, the conductance is not
  strongly dependent on the detailed dispersion relation of the leads.

\bibitem{note} The maximum conductance of the sample with only
periodic potential is $N_c {e^2 \over h}$, i. e. each conduction
channel may contribute a maximum of one unit conductance, since in
this case there is no coupling between (and therefore scattering into)
other channels.  Each injected electron either comes out on the other
side of the sample (if electron density and magnetic field $B$ are
such that the Fermi level is inside a subband) or is completely
reflected back (if Fermi level is in one of the gaps of the Hofstadter
butterfly).

\bibitem{Paulsson} Magnus Paulsson and Sven Stafstr\"om, Phys. Rev. B
{\bf 64}, 035416 (2001); {\em ibid} J. Phys. :Cond. Matt. {\bf 12},
9433 (2000).

\bibitem{Emberly}Eldon G. Emberly and George Kirczenow, Phys. Rev. B
{\bf 64}, 235412 (2001).

\bibitem{Onipko} Alexander Onipko, Yuri Klymenko, Lyuba Malysheva,
Phys. Rev. B {\bf 62}, 10480 (2000).

\bibitem{Kirkpatrick} D. J. Thouless and S. Kirkpatrick, J. Phys. C:
Solid State Phys., {\bf 14}, 235 (1981).

\bibitem{SuperLU} M. Baertschy, T. N. Rescigno,
W. A. Issacs, X. S. Li, and C. W. McCurdy, Phys. Rev. A {\bf 63}, 022712
(2001). For details of the software used see
http://www.nersc.gov/~xiaoye/SuperLU

\bibitem{Kirczenow}G. Kirczenow, Phys. Rev. B {\bf 39}, 10452 (1989).


\end{thebibliography}
\end{document}